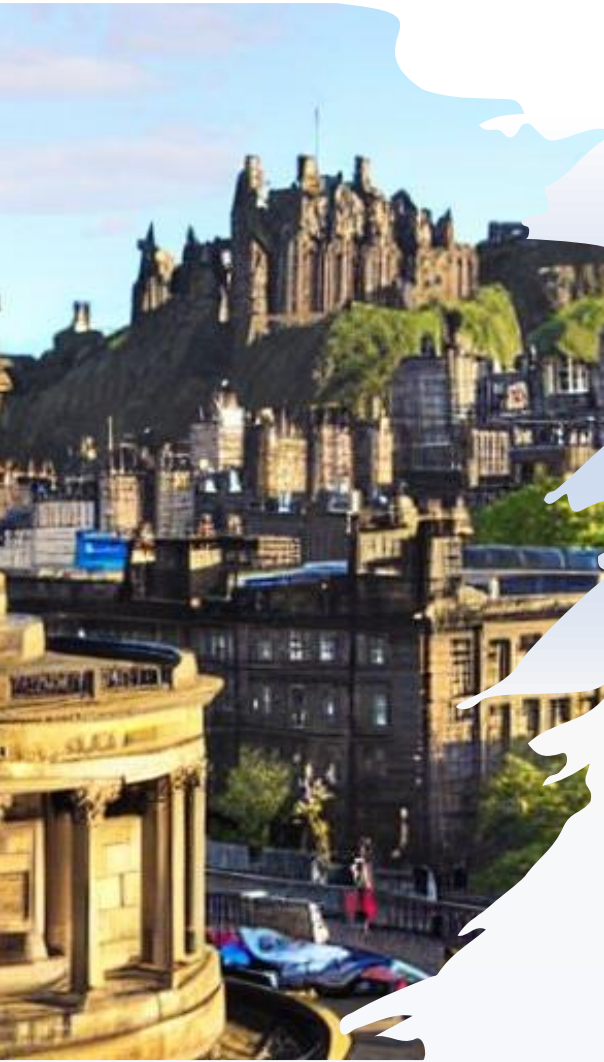

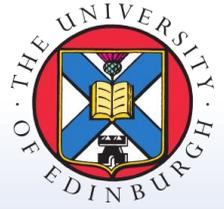

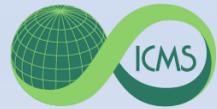

GELOCA
GRUPO ESPAÑOL DE LOCALIZACIÓN

EWG
EWGLA
LOCATIONAL
ANALYSIS

ICMS

EDINBURGH
MATHEMATICAL
SOCIETY

Glasgow
Mathematical
Journal
Trust

100
101
001
Seio
Sociedad
de Estadística
e Investigación
Operativa

UNIVERSIDAD DE LAS PALMAS
DE GRAN CANARIA

REDLOCA
RED DE LOCALIZACIÓN Y PROBLEMAS AFINES

GOBIERNO
DE ESPAÑA

MINISTERIO
DE ECONOMIA
Y COMPETITIVIDAD

# Proceedings of the XII International Workshop on Locational Analysis and Related Problems

7-8 September 2023
Edinburgh, Scotland

|  | Thursday Sep 7th |
| --- | --- |
| 09:30-09:45 | Opening Session |
| 09:45-10:45 | **Plenary Session I:** Antonio Rodríguez-Chía |
| 10:45-11:05 | Coffee break |
| 11:05-12:05 | **Session 1:** Hub Location and Applications |
| 12:10-13:10 | **Session 2:** Discrete Location |
| 13:10-14:10 | Lunch break |
| 14:10-15:10 | **Session 3:** Covering Problems |
| 15:15-16:15 | **Plenary Session II:** Paola Scaparra |
| 16:15-16:35 | Coffee break |
| 16:35-17:35 | **Session 4:** Continuous Location |
| 17:40-18:30 | Spanish Network Meeting |
| 18:30 | Social activity |

|  | Friday Sep 8th |
| --- | --- |
| 09:15-10:35 | **Session 5:** Applications |
| 10:35-10:55 | Coffee break |
| 10:55-11:55 | **Plenary Session III** Miguel Anjos |
| 12:00-13:00 | **Session 6:** Network Design |
| 13:00-14:00 | Lunch break |
| 14:00-15:20 | **Session 7:** Location and Routing |
| 15:20-15:40 | Coffee break |
| 15:40-16:40 | **Session 8:** Bilevel and Conic Programming |
| 17:45-19:15 | Ghostly Underground Tour |
| 19:30 | Workshop Dinner *Beirut (24 Nicholson Square)* |

# PROCEEDINGS OF THE XII INTERNATIONAL WORKSHOP ON LOCATIONAL ANALYSIS AND RELATED PROBLEMS (2023)

Edited by

Marta Baldomero-Naranjo

Víctor Blanco

Sergio García

Ricardo Gázquez

Joerg Kalcsics

Luisa I. Martínez-Merino

Juan M. Muñoz-Ocaña

Francisco Temprano

Alberto Torrejón

# Preface

The International Workshop on Locational Analysis and Related Problems will take place during September 7–8, 2023 in Edinburgh (United Kingdom). It is organized by the Spanish Location Network and the Location Group GELOCA from the Spanish Society of Statistics and Operations Research (SEIO). The Spanish Location Network is a group of more than 140 researchers from several Spanish universities organized into 7 thematic groups. The Network has been funded by the Spanish Government since 2003. The current project is RED2022-134149-T.

One of the main activities of the Network is a yearly meeting aimed at promoting the communication among its members and between them and other researchers, and to contribute to the development of the location field and related problems. As a proof of the internationalization of this research group, this will be the first time that the meeting is held out of Spain. The last meetings have taken place in:

- Elche (January 31– February 1, 2022)

- Sevilla (January 23–24, 2020)

- Cádiz (January 20–February 1, 2019)

- Segovia (September 27–29, 2017)

- Málaga (September 14–16, 2016)

- Barcelona (November 25–28, 2015)

- Sevilla (October 1–3, 2014)

- Torremolinos (Málaga, June 19–21, 2013)

- Granada (May 10–12, 2012)

- Las Palmas de Gran Canaria (February 2–5, 2011)

- Sevilla (February 1–3, 2010)

The topics of interest are location analysis and related problems. This includes location models, networks, transportation, logistics, exact and heuristic solution methods, and computational geometry, among others.

The organizing committee.

## Scientific committee:

- Maria Albareda Sambola (Universidad Politécnica de Cataluña)

- Víctor Blanco (Universidad de Granada)

- David Canca (Universidad de Sevilla)

- Sergio García Quiles (University of Edinburgh)

- Jörg Kalcsics (University of Edinburgh)

- Mercedes Landete (Universidad Miguel Hernández de Elche)

- Teresa Ortuño (Universidad Complutense de Madrid)

- Blas Pelegrín (Universidad de Murcia)

- Justo Puerto Albandoz (Universidad de Sevilla)

- Antonio M. Rodríguez-Chía (Universidad de Cádiz)

- Juan José Salazar (Universidad de La Laguna)

## Organizing committee:

- Sergio García Quiles (University of Edinburgh)

- Jörg Kalcsics (University of Edinburgh)

- Víctor Blanco (Universidad de Granada)

- Dolores R. Santos Peñate (Universidad de Las Palmas de G.C.)

- Gillian Kerr (ICMS)

- Lisa Garrett (ICMS)

# Contents

PROGRAM

# Thursday September 7th

*(Note that, in each session, the last speaker is the chair.)*

## 09:30-09:45 Opening Session

## 09:45-10:45 Plenary Session I:
##        Antonio M. Rodríguez-Chía

New formulations and solving methods for Single Allocation hub location problems

## 10:45-11:05 Coffee break

## 11:05-12:05 Session 1: Hub Location and Applications

Single-allocation hub location with upgraded connections

*M. Landete, J. M. Muñoz Ocaña, A. M. Rodríguez Chía and F. Saldanha-da-Gama*

The Waste-to-Biomethane Logistic Problem

*V. Blanco, Y. Hinojosa and V. Zavala*

The profit-oriented hub line location problem with elastic demand

*B. Cobeña, I. Contreras, L. I. Martínez-Merino and A. M. Rodríguez-Chía*

## 12:10-13:10 Session 2: Discrete Location

Budget-constrained center upgrading in the p-center location problem: A math-heuristic approach

*L. Anton-Sanchez, M. Landete and F. Saldanha-da-Gama*

The Induced Facility Location Problem with Upgrading

*I. Espejo and A. Marín*

Framework for Fairness Analysis in Location Problems

*I. Ljubić, M. A. Pozo, J. Puerto and A.Torrejón*

## 13:10-14:10 Lunch Break

## 14:10-15:10 Session 3: Covering Problems

A novel integrated approach for locating agricultural sensors based on the zoning and p-median problems

*T. Torres, V. Albornoz and R. Ortega*

Some insights on the Conditional Value-at-Risk approach for the Maximal Covering Location Problem with an uncertain number of facilities

*V. Blanco, R. Gázquez and F. Saldanha-da-Gama*

Setting second-level facilities in the multi-product maximal covering location problem

*M. Baldomero-Naranjo, L.I. Martínez-Merino and A. M. Rodríguez-Chia*

## 15:15-16:15 Plenary Session II: Paola Scaparra

Unlocking sustainable solutions: Operations Research for Global Goals

## 16:15-16:35 Coffee break

## 16:35-17:35 Session 4: Continuous Location

Planar location problems with uncertainty demand points: robustness concepts

*J.A. Mesa and A. Schöbel*

Touring polytopes, the weighted region and some related location problems

*J. Puerto, M. Labbé and M. Rodríguez*

Exact Solution Methods for the $p$-Median Problem with Manhattan Distance and a River with Crossings

*T. Byrne and A. Suzuki*

## 17:40-18:30 Spanish Network on Location Analysis Meeting

## 18:30 Social Activity

# Friday September 8th

*(Note that, in each session, the last speaker is the chair.)*

## 09:15-10:35 Session 5: Applications

An exact approach towards solving a hydrogen refuelling network design problem with pipelines

*K. Searle, C. Searle and J. Kalcsics*

Mixed Integer Programming Models for Installation of Offshore Wind Turbines

*K. Akartunali, M. Doostmohammadi and C. E. Onyi*

The Pipelines and Cable Trays Location Problem in Naval Design

*V. Blanco, G. González-Domínguez, Y. Hinojosa, D. Ponce, M. A. Pozo and J. Puerto*

Heuristic segmentations in Electron Tomography using the DOMP

*J. M. Muñoz-Ocaña, A. M. Rodríguez-Chía and J. Puerto*

## 10:35-10:55 Coffee break

## 10:55-11:55 Plenary Session III: Miguel Anjos

Optimal Location of Electric Vehicle Charging Stations

## 12:00-13:00 Session 6: Network Design

Connected graph partitioning with aggregated and non-aggregated gap objective functions

*E. Fernandez, I. Lari, J. Puerto, F. Ricca and A. Scozzari*

Budget constrained cut problems

*J. Puerto and J. L. Sainz-Pardo*

New Formulations and Column Generation Algorithm of the Minimum Normalized Cuts Problem

*D. Ponce, J. Puerto and F. Temprano*

## 13:00-14:00 Lunch break

## 14:00-15:20 Session 7: Location and Routing

The Min Max Multi-Trip Location Arc Routing Problem
*T. Corberán, I. Plana and J. M. Sanchis*

Prize-collecting Location Routing on Capacitated Trees
*E. Fernandez and M. Munoz-Marquez*

The load-dependent drone general routing problem
*I. Plana. J. M. Sanchis and P. Segura*

The Hampered Travelling Salesman Problem with Neighbourhoods
*J. Puerto and C. Valverde*

## 15:20-15:40 Coffee break

## 15:40-16:40 Session 8: Bilevel and Conic Programming

Combinatorics of Extended Representations of generalized power cones
*V. Blanco and M. Martínez-Antón*

A bi-level facility delocation problem with spatially loyal customers: exact methods and metaheuristics
*J. F. Camacho-Vallejo, J. C. García-Vélez, D. Ruiz-Hernández and J. A Díaz*

Bi-Level Service Design Problem considering outsourcing costs
*E. Fernandez, I. Ljubić and N. Zerega*

## 17:45-19:15 Ghostly Underground Tour

## 19:30 Workshop Dinner at Beirut (24 Nicholson Square)

INVITED SPEAKERS

# Optimal Location of Electric Vehicle Charging Stations

Miguel Anjos [1]

[1]*University of Edinburgh, The United Kingdom*,   miguel.f.anjos@ed.ac.uk

The increase of electric vehicle (EV) adoption in recent years has correspondingly increased the importance of providing adequate charging infrastructure for EV users. For a charging service provider, the fundamental question is to determine the optimal location and sizing of charging stations with respect to a given objective and subject to budget and other practical constraints. Practical objectives include maximizing EV adoption as part of a public policy on electric transportation, and maximizing the profit gained from providing this service, in which case the price of charging may also be optimized. In this talk, we will present an overview of our work in this area and discuss open directions for future research.

# New formulations and solving methods for Single Allocation hub location problems

Antonio M. Rodríguez-Chía [1]

[1] *Departamento de Estadística e Investigación Operativa, Universidad de Cádiz, Spain,*
antonio.rodriguezchia@uca.es

The most promising formulations and solution methods in the literature for solving the single-allocation hub location problem (SAHLP) are revised. Then, a new compact formulation for SAHLP with fewer variables than the previous Integer Linear Programming formulations in the literature is introduced (valid also when costs costs are not based on distances and not satisfying triangle inequality. Moreover, costs can be given in aggregated or disaggregated way. Different families of valid inequalities that strengthen the formulation are developed. A branch-and-cut algorithm based on a relaxed version of the formulation is designed, whose restrictions are inserted in a cut generation procedure together with two sets of valid inequalities. The performance of the proposed methodology is tested on well-known hub location data sets and compared to the most recent and efficient exact algorithms for single-allocation hub location problems. Extensive computational results prove the efficiency of our methodology, that solves large-scale instances in very competitive times.

# Unlocking sustainable solutions: Operations Research for the Global Goals

Paola Scaparra [1]

[1] *University of Kent, The United Kingdom,*  m.p.scaparra@kent.ac.uk

The United Nations Sustainable Development Goals (SDGs) are a universal call to action to end poverty, protect the planet, and improve the lives and prospects of everyone, everywhere. Operational Research has a major role to play in finding solutions to sustainable development challenges and helping the achievement of the SDGs. This talk discusses two projects, funded by the UK Global Challenges Research Fund (GCRF), to build more resilient, inclusive, and sustainable communities in Southeast Asia. The first project (GCRF-OSIRIS) used OR tools integrated with models and output from the fields of meteorology, hydrology, and social science to minimise the impacts of severe flooding in urban areas of Vietnam. Because flooding disproportionally impacts the lives of women and poor communities, the views and concerns of these vulnerable groups were elicited and taken into account during the model building phase to identify community-responsive flood mitigation solutions. Building on the success of GCRF-OSIRIS, a follow-up project (CREST-OR) focused on fostering self-sustained OR communities in Cambodia, Indonesia, Laos, Myanmar, and Vietnam by equipping future generations with the relevant OR expertise needed to tackle a range of sustainable development challenges. The talk will conclude by highlighting additional opportunities for OR and location analysis researchers to help realise a sustainable future in developing countries and accelerate progress of the SDG agenda.

ABSTRACTS

# Mixed integer programming models for installation of offshore wind turbines

Kerem Akartunali[1], Mahdi Doostmohammadi[1], Charles E. Onyi[1]

[1]*Department of Management Science, University of Strathclyde, Glasgow G4 0GE, UK.*
kerem.akartunali@strath.ac.uk, m.doostmohammadi@strath.ac.uk, charles.onyi@strath.ac.uk

With millions of pounds involved in the construction of offshore wind turbines, there is need to optimize the whole process as to reduce the cost and duration of the final process involved in wind energy generation. Specifically, this project focuses on the derivation of mixed integer linear programming formulations for the installation procedure of offshore wind turbines and then uses exact methods to solve those models. However, we devise a shortest path-like formulations using strong multicommodity flow models of the network. We propose a polyhedral analysis of the set of solutions by identifying novel problem-specific valid inequalities require in the characterization of the convex hull of feasible solutions. Also we design an efficient cutting plane algorithm that identifies the optimal configuration of vessel schedules in order to minimize the installation duration and cost. We present preliminary computational results on the problem specific and the literature instances that are related to the problem formulation. Our method efficiency shall be tested and compared to most of the generated instances in the literature and real word data as well as those related to the subject matter.

# Budget-constrained center upgrading in the $p$-center location problem: A math-heuristic approach [*]

Laura Anton-Sanchez,[1] Mercedes Landete,[1] and Francisco Saldanha-da-Gama[2]

[1]*Departamento de Estadística, Matemáticas e Informática, Centro de Investigación Operativa, Universidad Miguel Hernández, Spain,*   l.anton@umh.es     landete@umh.es

[2]*Sheffield University Management School, Sheffield, United Kingdom,* francisco.saldanha-da-gama@sheffield.ac.uk

## 1. Introduction

Recently, we have investigated different upgrading strategies in the context of the $p$-center problem [1]. We focus on the so-called *unweighted vertex-restricted p-center problem* and consider the possibility of upgrading a set of connections to different centers as well as the possibility of upgrading entire centers, i.e., all connections made to them. In the aforementioned work, two variants for these perspectives are analyzed: in the first, there is a limit on the number of connections or centers that can be upgraded; in the second, an existing budget is assumed for the same purpose. We introduce different mixed-integer linear programming models for those problems (based on those previously proposed by Daskin [3] and Calik and Tansel [2]) and we show that, in most cases, an optimal solution can be obtained within an acceptable computing time using an off-the-shelf solver. Nevertheless, this is not the case for the budget-constrained center-upgrading model, i.e., when an exogenous budget is considered for upgrading centers. The aim of the work at hand is to present the development of a math-heuristic seeking high-quality feasible solutions in that specific case.

---

[*]This work was partially supported by the grants PID2021-122344NB-I00, PID2019-105952GB-I00/ AEI /10.13039/ 501100011033, and PGC2018-099428-B-100 by the Spanish Ministry of Science and Innovation, PROMETEO/2021/063 by the governments of Spain and the Valencian Community, project CIGE/2021/161 by the Valencian Community, and UIDB/04561/2020 by National Funding from FCT — Fundação para a Ciência e Tecnologia, Portugal.

# 2. Math-heuristic procedure for budget-constrained center upgrading

We propose a genetic algorithm for budget-constrained center upgrading, whose structure is as follows: once an initial population has been generated, crossover, mutation, and local search operators are carried out in all iterations until the stopping condition occurs. Furthermore, an intensified local search is performed in some iterations, which is ruled by some given *probability*. We have encoded the solutions in such a way that we only save the value of the $p$ open centers. Hence, each individual is a combination of $p$ centers from the candidates. We approximate the fitness of an individual (objective function value) by using a linear optimization model that provides a good upper bound on that value. Moreover, we try to improve this upper bound further by reallocating some demand nodes. As a consequence of the way the fitness of an individual is obtained, our genetic algorithm turns out to be a math-heuristic, in which the calculation of the fitness function is not trivial.

The results show that the math-heuristic finds good feasible solutions in a short time. For all instances, the results provided by the algorithm, even if they are not optimal solutions, provide hopefully good upper bounds for the problems.

# Setting second-level facilities in the multi-product maximal covering location problem

Marta Baldomero-Naranjo,[1] Luisa I. Martínez-Merino,[2]
Antonio M. Rodríguez-Chía[2]

[1]*Universidad Complutense de Madrid, Spain,*  martbald@ucm.es

[2]*Universidad de Cádiz, Spain,*  luisa.martinez@uca.es antonio.rodriguezchia@uca.es

In hierarchical facility location problems, the goal is to locate a set of interacting facilities at different levels of a hierarchical framework. In this context, we introduce a model which considers a first-level system of already established services (factories, product sources, etc.), a second-level system of facilities (warehouses, shops, etc.) to determine their location and their supply of products, and a third-level system of clients demanding different products produced in the first-level and provided by the second-level facilities.

Each client demands different products and has distinct preferences for the same product depending on the first-level facility producing that product. In this model, called multi-product maximal covering second-level facility location problem, there is a maximum number of different products that can be offered at each second-level facility and also a budget constraint for the total cost of the facility locations.

The aim is to locate a set of second-level facilities and to decide the size of them (number of offered products), in such a way that the covered clients' demand is maximized. Therefore, in order to satisfy a customer's demand there must be a double coverage, the customer must be covered by a second-level facility, and this, in turn, by a first-level facility.

We propose a mixed integer linear program (MILP) for this problem which is reinforced by the use of valid inequalities. For cases where the number of valid inequalities is exponential, different separation procedures are developed. In addition, three variants of a heuristic algorithm are pro-posed. An extensive computational analysis is carried out. This illustrates the usefulness of the valid inequalities and the corresponding separation methods, as well as, shows the good performance of the proposed heuristic algorithms.

# Some insights on the Conditional Value-at-Risk approach for the Maximal Covering Location Problem with an uncertain number of facilities[*]

Víctor Blanco,[1] Ricardo Gázquez,[2] and Francisco-Saldanha-da-Gama[3]

[1] *Institute of Mathematics (IMAG), Universidad de Granada, Spain,*   vblanco@ugr.es

[2] *Department of Statistics, Universidad Carlos III de Madrid, Spain,*   rgazquez@ugr.es

[3]*Sheffield University Management School, Sheffield, United Kingdom,*
francisco.saldanha.da.gama@sheffield.ac.uk

The maximal covering location problem (MCLP) lies in the area of Covering Problems which is a well-defined branch of Location Science [2]. In particular, the MCLP maximize the amount of demand covered within a maximal service distance by locating a fixed number of facilities [1].

Many different approaches to the problem have been developed since its introduction, among them the MCLP with uncertainty. Uncertainty is usually assumed in aspects such as travel time (which would affect distance) or customer demands. Very few papers have worked with uncertainty in the number of facilities to be located (see e.g., [3,4]).

In this talk, we focus in the MCLP with uncertainty on the number of facilities to be located and treat it as a two-stage problem. In the first stage, a decision has to be made on an initial set of facilities to locate; it is a here-and-now decision. In the second stage, a set of additional locations is selected, and we assume that the exact number of facilities to locate in the second stage is not known in advance—it corresponds to information that will be revealed at a future point in time, for example, based on a budget whose volume is not yet known.

[*]Research partially supported by the Spanish Ministry of Science and Innovation through project RED2022-134149-T

This problem is useful in many applications such as telecommunication antennas and repeaters, in managing the location of cameras over a city and even in problems of electric vehicle charging stations, where cities are currently developing plans for the placement of the stations.

Finally, we model the problem using the Conditional Value-at-Risk (CVaR). This consists of minimizing a convex combination of the maximum regret with respect to a set of scenarios with jointly probability at least $\alpha$ and the expected regret with respect to the remainder scenarios (see also, [5,6])

We derive a mathematical programming model for the problem and derive some interesting conclusions on the obtained results, both on synthetic and real-world datasets.

# The Pipelines and Cable Trays Location Problem in Naval Design

V. Blanco,[1] G. González,[2] Y. Hinojosa,[3] D. Ponce,[4] M. A. Pozo [5] and J. Puerto.[6]

[1]*Granada University, Spain,*   vblanco@ugr.es

[2]*Granada University, Spain,*   emaildegabri@gmail.com

[3]*Seville University, Spain,*   yhinojos@us.es

[4]*Seville University, Spain,*   dponce@us.es

[5]*Seville University, Spain,*   miguelpozo@us.es

[6]*Seville University, Spain,*   puerto@us.es

This paper focuses on determining optimal locations for pipelines and cable trays in naval design, aiming to minimize a user-defined cost function. The problem involves finding the right number and types of cable tray routes between various devices. We simplify the problem by reducing it to an *ad hoc* min-cost multicommodity flow problem. Discretizing the entire space into a graph structure helps address the problem mathematically. In cases where there are no additional constraints and only a single pipeline or cable tray needs routing, finding the optimal path between two nodes in a graph can solve the problem efficiently using Dijkstra's algorithm [1]. However, practical scenarios often involve routing multiple pipelines and cable trays together, requiring additional constraints imposed by technical requirements. These constraints serve the following purposes: a) Ensuring minimum allowed distances between services of different types based on pipeline or cable tray width. b) Guaranteeing a minimum allowable distance between two elbows on the same pipe or cable tray. c) Limiting the capacity of a cable tray to a specific value. This paper enhances the existing model described in [2] by integrating cable trays, introducing complexity by considering pipeline paths and tray placement decisions. Optimal solutions minimize the overall occupied area by pipelines and cable trays.

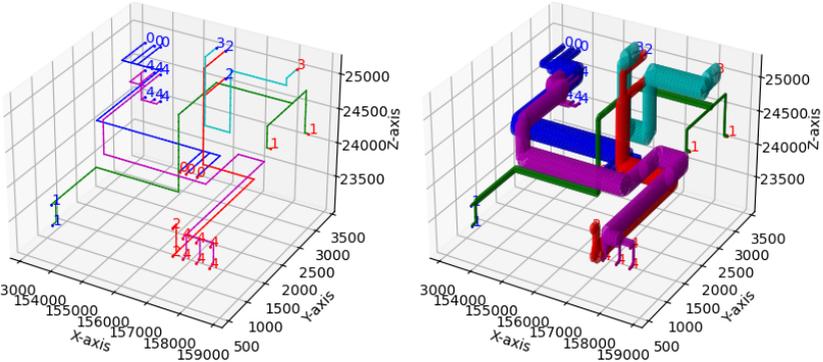

*Figure 1.* Graphical display of the solution for the case study as a collection of edges (left) and depicting actual pipelines and trays (right).

To handle the large number of constraints, we adopt a two-step approach. Initially, a relaxed formulation is solved without constraints, and if violations occur, the violated constraints are added to a constraint pool. The problem is then solved again, incorporating the additional constraints, until a feasible solution satisfying all constraints is obtained. The procedure is integrated into a branch-and-cut scheme using callbacks available in commercial off-the-shelf software such as Gurobi, CPLEX, or FICO. For larger instances, we propose a family of heuristic algorithms with two phases: I) Constructing initial cable tray paths, and II) Transforming these paths into feasible cable trays that meet technical requirements. Different strategies within each phase lead to several algorithms, which are compared through computational experiments using two types of instances: randomly generated instances and instances with well-defined corridors to assess the effectiveness of our methodology. Finally, we analyze a real-life case study provided by Ghenova, a leading Naval Engineering company, to validate the effectiveness of our proposed approach. The representation of the obtained solution for the case study is presented in Figure 1.

# The Waste-to-Biomethane Logistic Problem
*

Víctor Blanco[1], Yolanda Hinojosa[2], and Víctor Zavala[3]

[1]*Institute of Mathematics (IMAG), Universidad de Granada, Spain*  vblanco@ugr.es

[2]*Institute of Mathematics (IMUS), Universidad de Sevilla, Spain*  yhinojos@us.es

[3]*Chemical and Biological Engineering, University of Wisconsin-Madison, United States*
zavalatejeda@wisc.edu

The Green Deal was an initiative proposed by the European Community in 2019 as an statement of intents on the roadmap of Europe to implement the United Nation's 2030 and 2050 Agendas for Sustainable Development designed to mitigate the effect of the climate change. One of challenges of the communication is to decarbonize the energy system by developing a new power sector based on renewable sources. It is a fact that one of the main tools to achieve the proposed goals is the use of biogas as an alternative renewable energy source to carbon-based energies, since it contributes the reduction of greenhouse gases but also to the development of the circular economy through the anaerobic digestion of organic waste from different sources and its transformation into fuel. Since biomethane is the same molecule as natural gas, it can be distributed via the existing gas distribution networks, facilitating the transition from natural gas to biogas energy. Thus, many countries are making special interest in the assessment of the installation and adaptation of biogas plants and the evaluation of the amount of energy that these plants are capable to produce. The main draw-backs of is that the implementation of efficient biogas plants as an direct economic effect in consumers and businesses, and thus, in order to implement and efficient and sustainable biogas distribution system one has to adequately design an affordable and robust logistic plan for the different manure, biomethane, waste, etc involved in the process, making alternative energy system affordable and effective. In this phase, it is crucial to decide where to locate the biogas plants (or the transhipments plants to distribute the products), how to collect the manures from the farms, fields, etc to those plants, how to connect (if needed) the different types of plants, how to distribute the final biomethane to the gas distribution network and the possible wastes to either dumps or for fertilizer to some of the users. At this point, mathematical optimization plays a very important role since it is the main tool to determine solutions to complex logistics system as the one here [2].

---

*The authors of this research acknowledge financial support by the Spanish Ministerio de Ciencia y Tecnologia, Agencia Estatal de Investigacion and Fondos Europeos de Desarrollo Regional (FEDER) via project PID2020-114594GB-C21. The authors also acknowledge partial support from projects FEDER-US-1256951, Junta de Andalucía P18-FR-1422, P18-FR-2369, B-FQM-322-UGR20, NetmeetData: Ayudas Fundación BBVA a equipos de investigación científica 2019, and the IMAG-Maria de Maeztu grant CEX2020-001105-M /AEI /10.13039/501100011033. The first author also acknowledges the financial support of the European Union-Next GenerationEU through the program "Ayudas para la Recualificación del Sistema Universitario Español 2021-2023"

In this work we study the Waste-to-Biomethane Logistic Problem, that consists of deciding the location of the different plants and pipelines involved in the transformation and distribution of biomethane, and the way the product is transported, minimizing the overall transportation cost with the given installation budget. This logistic system behind the biogas production is complex, with different agents, different production and conversion technologies, and different types of demand centers [1]. We provide a general and flexible family of mathematical optimization models to make decisions in the framework of locating biogas plants and distributing the different types of element along a complex network.

# Combinatorics of Extended Representation of generalized power cones

Víctor Blanco[1] and Miguel Martínez-Antón[1]

[1]*Institute of Mathematics (IMAG), Universidad de Granada, Spain*    vblanco@ugr.es; mmanton@ugr.es

The use of conic structures in mathematical optimization problems has been widely recognized within the last years. The development of interior points methods with specialized barriers have given room to efficiently solve problems involving cones with high practical interest. The incorporation of exact techniques to solve problems involving cones in different off-the-shelf optimization solvers has allowed practitioners to use these tools to make decisions in different fields. This is the case of continuous facility location [2], machine learning [3], power flow [4], radiotherapy [11], portfolio selection [6], geometric programming [1,9], among many others. Thus, a lot of theoretical results have emerged to understand the geometry of different representations of cones [5,7,10].

In this paper we analyze a general type of power cone, the $(\boldsymbol{\alpha}, p)$-order cone, which is defined as:

$$\mathcal{K}_p^{d_1+d_2}(\boldsymbol{\alpha}) = \{(\mathbf{x}, \mathbf{z}) \in \mathbb{R}^{d_1} \times \mathbb{R}_+^{d_2} : \|\mathbf{x}\|_p \leq \mathbf{z}^{\boldsymbol{\alpha}}\},$$

for $p \in \mathbb{R}_+ \cup \{\infty\}$ ($p \geq 1$), $d_1, d_2 \in \mathbb{Z}_+$ ($d_2 \geq 1$), and $\boldsymbol{\alpha} \in \{(\alpha_1, \ldots, \alpha_{d_2}) \in \mathbb{R}^+ : \sum_{j=1}^{d_2} \alpha_j = 1\}$. This cone has as particular cases the $p$-order cone and the power cone, and its efficient representation into an optimization problem allow to solve tons of problems that appear in different fields, particularly in continuous location problems.

An extended representation of a set is an equivalent reformulation of the set through the inclusion of the set in a higher dimensional space and its after-transformation as a product of simpler sets. In this paper we derive different extended representations of any $(\boldsymbol{\alpha}, p)$-power cone by means of simpler power cones and study its complexity. In case $\boldsymbol{\alpha}$ is a rational vector and $p$ is a rational number, we provide extended representations of the cone as 3-dimensional rotated quadratic cones. This topic has been addressed for $\boldsymbol{\alpha} = \left(\frac{r_1}{2^m}, \frac{r_2}{2^m}, \frac{r_3}{2^m}\right)$ in [8]; and for $\boldsymbol{\alpha} = \left(\frac{r_1}{2^m}, \ldots, \frac{r_n}{2^m}\right)$ where $m \in \mathbb{N}$ in [5]. The main advantage of these simpler cones is that they are efficiently handled by most of the off-the-shelf optimization solvers. We show that a naive representation is highly inefficient compared to a guided representation, and provide bounds for the complexity of the representations in terms of $\alpha$ and $p$.

# Exact Solution Methods for the p-Median Problem with Manhattan Distance and a River with Crossings

Thomas Byrne[1] and Atsuo Suzuki[2]

[1]*University of Strathclyde, United Kingdom,*  tom.byrne@strath.ac.uk

[2]*Nanzan University, Japan,*  atsuo@nanzan-u.ac.jp

Rivers are a source of fresh, drinkable water, carrying and distributing important salts and nutrients. In supporting plant and animal life, rivers bestow a useful food source as well allow the use of irrigation, essential to food production. Vegetation that thrives around rivers facilitates lower air and surface temperatures by releasing moisture into the atmosphere and providing shade. Additionally, rivers are often crucial for transportation and commerce and more recently have been developed to produce electricity and to provide for leisure activities.

Not surprisingly, large population densities congregate around large rivers. While humans inhabit around 38% of the world's surface area, humans inhabit over 40% of the area for which a river is the closest water feature. In fact, on average, the closest body of water to a human is a large river at a median distance of 2.2km [3].

However, in current facility location models, the pivotal influence on travel that a river exerts has not yet been adequately addressed. Despite providing such vital resources to a settlement to this day, rivers are usually impassable but for pre-constructed crossings and so present a very real barrier to anyone hoping to travel besides or to the other side of the river.

We propose a collection of exact solution methods for the $p$-median problem with Manhattan ($l_1$) distance and a river, a problem we shall refer to as pMPL1($R,C$) (where $R$ and $C$ represent the river and its crossing respectively). pMPL1($R,C$) requires the locating of several facilities in order to minimise the total travelling distance from the given demand points to their nearest facility. For any route across the river, a crossing point is cho-sen from a given set of crossings so as to minimise the shortest paths where distance is measured using the Manhattan metric which, taking into account the grid structure prevalent in many cities' transport infrastructure, is most representative when considering urban applications of facility location.

It is known that the 1-median problem with Manhattan distance is easily solved because its objective function is a piecewise linear function. For the $p$-median problem with Manhattan distance, a few heuristic algorithms have already been proposed. However, even for this well-known problem, effective algorithms to obtain its exact solution have not yet been studied [1]. The objective function of pMPL1($R$,$C$) is not convex and there are many local minima. Resorting to the existing heuristic methods risks obtaining not the exact solution but one of these local minima. A naïve solution method is to enumerate all the candidate points and to evaluate the objective function at each in order to find the solution. This, not surprisingly, is time-consuming. Instead, we construct a BTST algorithm [2] for pMPL1($R$,$C$) which obtains the exact solution in practical computational time. We compare our algorithm with the naïve enumeration method and show the effectiveness of our algorithm.

# A bi-level facility delocation problem with spatially loyal customers: exact methods and metaheuristics

José-Fernando Camacho-Vallejo[1], Juan-Carlos García-Vélez[2], Diego Ruiz-Hernández[3], and Juan A. Díaz[4]

[1]*Tecnologico de Monterrey, Escuela de Ingenieria y Ciencias, Mexico,* fernando.camacho@tec.mx

[2]*Facultad de Ciencias Físico-Matemáticas, Universidad Autónoma de Nuevo León, México,* jgarciav@uanl.edu.mx

[3]*Sheffield University Management School, Sheffield, UK,* d.ruiz-hernandez@sheffield.ac.uk

[4]*Departamento de Física, Actuaría y Matemáticas, Universidad de las Américas Puebla, México,* juana.diaz@udlap.mx

The aim of this study is to contribute to the literature of competitive delocation by addressing the problem faced by a large firm (the leader) who tries to reduce its network by closing a predefined number of facilities, on the knowledge that his competitor (the follower) -who is also engaged in a restructuring process- will react strategically to his actions. An important contribution of this study is the introduction of a novel notion of loyalty: the *loyalty radius*. This notion, presumes that certain customers will still be willing to visit a facility of their favorite provider, despite the fact that a competitor's facility may be located at a shorter distance, provided that the open facility is located within certain admissible distance.

The problem is modelled as a bi-level competitive delocation model and is formulated as a variant of the well-known $(r|p)$-centroid problem. These type of problems are complex to solve. For instance, even in the simplest form where both levels are linear programs, [1] demonstrates that the bi-level problem is NP-hard. In our case, we consider a binary-binary linear bi-level problem.

Two exact methods are used: a proposed depth-first search implicit enumeration algorithm based on the ideas given in [2] and a generic solver of mixed-integer linear bi-level programming problems presented in [3]. Additionally, two metaheuristics are developed: a nested GRASP algorithm and a nested hybrid GRASP-tabu search algorithm.

The four algorithms are used for solving a diverse set of instances of the problem. The analysis of an instance based on a real-life-setting, and intensive numerical experimentation, provide evidence on the effectiveness of the algorithms for the solution of real-size problems, and confirm the practical validity of our modelling approach. Our results also highlight the importance of taking into account spatial loyalty in network restructuring scenarios. It is noteworthy to emphasize that both metaheuristic algorithm obtain the optimal solution of the problem in less computational time. Its effectiveness and efficiency has been proven by the extensive computional experimentation conducted.

The numerical results highlight the importance of taking into consideration customers' behaviour, in particular loyalty, when engaging in restructuring processes. The main conclusion is that whereas in absence of loyalty restructuring efforts tend to benefit the leader, the existence of loyalty changes the balance, and the follower seems to be able to capture a larger segment of the market. Additionally, our results suggest that acknowledging the existence of loyalty in the planning process tends to bring larger benefits to the follower, who seems to be able to better adapt his/her restructuring strategy.

# The profit-oriented hub line location problem with elastic demand

Brenda Cobeña,[1] Ivan Contreras[1], Luisa I. Martínez-Merino[2], and Antonio M. Rodríguez-Chía[2]

[1]*Concordia University and Interuniversity Research Centre on Enterprise Networks, Logistics, and Transportation, Canada,*
brenda.cobena@mail.concordia.ca, ivan.contreras@concordia.ca

[2]*Universidad de Cádiz, Spain,*  luisa.martinez@uca.es, antonio.rodriguezchia@uca.es

The hub line location problem (HLLP) was introduced in [1] and it has relevant applications in public transportation planning design (tram, subways, express bus lane, etc.) and road network design. The aim of the HLLP is to locate $p$ hubs connected by a path that is composed by p-1 hub arcs in such a way that the total travel time of the origin-destination pairs is minimized. If the direct travel time between an origin-destination is smaller than any path using the line, the demand is routed without using the hub line. HLLP assumes that demand between each origin and destination can be a priori estimated and that the resulting constructed hub line network does not have any effect on the demand.

In this work, we extend the HLLP. Particularly, we introduce the profit-oriented hub line location problem with elastic demand (ED-HLLP). This model uses a gravity model to include the elasticity of demand. Gravity models have been previously used to model elastic demand in network design problems, see [2]. The main goal of ED-HLLP is to maximize the revenue of the total time reduction provided by the hub line considering elastic demand. Thus, this new model takes into account the long-term impact on the demands that opening a hub line can have.

We propose different non-linear formulations to address this problem. The main difference between these formulations is the way in which the line structure is modeled. Besides, we introduce three main linear formulations. These linear formulations use the possible paths using the hub line as variables. Consequently, it is necessary to use a preprocessing phase that calculates the candidate paths for each origin-destination. For this preprocessing phase, we provide efficient algorithms to create all possible candidate paths. Finally, we also present a computational experience that evaluates the strengths and limits of these formulations for ED-HLLP.

# The Min Max Multi-Trip Location Arc Routing Problem

Teresa Corberán[1], Isaac Plana [2] and José María Sanchis [3]

[1]*Universitat de València, Spain,*   tecorfa@alumni.uv.es

[2]*Universitat de València, Spain,*   isaac.plana@uv.es

[3]*Universitat Politècnica de València, Spain,*   jmsanchis@mat.upv.es

Unlike ground vehicle routing problems, drone routing problems are characterized by the fact that drones can enter and leave the edges at any point and serve only part of them. Therefore, the service of an edge can be shared by several drones, making an already difficult problem much harder. As in previous works, we approach the problem by adding a set of intermediate points to each original edge, obtaining then a classic arc routing problem where a set of given edges must be traversed. In the Min Max Multi-Trip Location Arc Routing Problem (MM-MT-LARP), we consider a depot from which a set of P trucks, each one carrying a drone, must travel to P out of D available points ($D \geq P$), where the drone is launched. Each drone has a limited autonomy which allows it to fly a maximum time L before having to get back to the launching point to change its battery so that it can start another route. Once the drone has completed all its routes, the truck goes back to the depot. The goal of the MM-MT-LARP is to determine the launching point for each truck and find a set of drone routes for each truck, each one starting and ending at its launching point and with flight time not greater than L, in such a way that all the drones' routes jointly traverse all the given edges and the largest total time of all the trucks (time of traveling to the launching point, flight time of the drone and time of traveling back to the depot) is minimized. In this talk, we present an integer linear programming formulation and a metaheuristic algorithm for the MM-MT-LARP, as well as some preliminary computational results on a set of instances with different characteristics.

# The Induced Facility Location Problem with Upgrading

Inmaculada Espejo,[1] and Alfredo Marín[2]

[1]*Departamento de Estadística e Investigación Operativa, Universidad de Cádiz, Spain,* inmaculada.espejo@uca.es

[2]*Departamento de Estadística e Investigación Operativa, Universidad de Murcia, Spain,* amarin@um.es

In a classical facility location problem, a graph with a weight (cost) on each edge of the graph is considered. The goal is finding optimal locations for a set of facilities on the nodes of a graph with respect to some objective function and allocate the demand points to the facilities.

We consider a new model for facility location problems on graphs where two kinds of weights are associated with each edge, namely allocation and derived costs. The allocation cost is related to the cost of assigning the demand points to the facilities. The derived cost represents any cost derived of the allocation of the demand points. Each customer will be assigned to the facility providing the lowest allocation cost. Moreover, a budget is given to reduce the derived costs. The aim is to simultaneously find the location of facilities and the distribution of a budget in the edges of the graph to reduce the derived costs associated to the allocation of the demand nodes to the facilities, in order to minimize the upgraded derived costs.

Some authors have already applied upgrading approach to several location problems. Related to median problems, [3, 6, 11, 12], and regarding the center problems, [8, 9, 13]. The 1-centdian problem was studied in [10] and the obnoxious median location problems on trees have been studied in [1, 2, 7]. The upgrading version of the maximal covering location problem and the hub-location problems have been considered in [4] and [5], respectively.

To the best of our knowledge, facility location problems considering arc upgrading at the same time and with two kinds of weights associated with the arcs of a network have not been studied yet.

# Connected graph partitioning with aggregated and non-aggregated gap objective functions[*]

Elena Fernández[1], Isabella Lari[2], Justo Puerto[3], Federica Ricca[2], Andrea Scozzari[4]

[1] *University of Cádiz, Spain,* elena.fernandez@uca.es

[2] *Sapienza Università di Roma, Italy,* {isabella.lari, federica.ricca}@uniroma1.it

[3] *Institute of Mathematics University of Seville (IMUS) and Universidad de Sevilla, Spain,* puerto@us.es

[4] *Università degli Studi Niccolò Cusano Roma, Italy,* andrea.scozzari@unicusano.it

This work deals with the problem of partitioning a graph into $p$ connected components by optimizing some balancing objective functions related to the vertex weights. The problem of partitioning a graph into a given number of connected components is a long studied topic in the literature [1]. It arises in a wide variety of practical applications in different contexts, such as engineering, economics, logistics, psychology, medicine, and electoral systems [2–5]. A connected graph $G$ is given, together with an integer value $p$ corresponding to the number of connected components of the partition. Weights are assigned to the elements of the graph (vertices or edges) and the aim is to partition the graph into $p$ connected components that are *balanced* (or *uniform* or *homogeneous*) with respect to such weights, according to a given criterion.

[*]Partially supported by the Agencia Estatal de Investigacion and Fondos Europeos de Desarrollo Regional (FEDER) via projects by MINECO MTM2019-105824GB-I00, PID2020-114594GB-C21, FEDER-US-1256951, Junta de Andalucía P18-FR-422, CEI-3-FQM331, FQM-331, and NetmeetData: Ayudas Fundación BBVA a equipos de investigación científica 2019. The authors acknowledge Sapienza Università di Roma for the financial support of the visiting period of Elena Fernández and Justo Puerto at the Dept. MEMOTEF (C26V16M9NH, C26V20SPJ5).

We consider objective functions based on the *gap* or *range* of the partition's components, i.e., the difference between the maximum and minimum weight of a vertex in the component. In particular, we define the notion of *aggregated gap* as the sum of the differences between the weights of the vertices and the minimum weight of a vertex in the component. Based on this notion, we introduce two new optimization problems, the Min-Sum Aggregated Gap Partition (MSAGP) and the Min-Max Aggregated Gap Partition (MMAGP), where the objective function considers aggregated gaps. To the best of our knowledge, they have not been investigated yet for connected $p$-partition problems. We study their complexity, proving that these problems are NP-hard on general graphs.

We provide several mathematical formulations for the set of constraints of MSAGP and MMAGP, adopting flow-based constraints for modeling connectivity in a partition. These formulations are quite general, and can be easily adapted to other partitioning problems studied in the literature like Min-Sum Gap Partitioning (MSGP) or the Min-Max Gap Partitioning (MMGP). The use of ad-hoc symmetry breaking constraints allows applying a variable fixing scheme notably facilitating the optimization of the problem, for MSAGP and MMAGP, as well as for MSGP and MMGP.

We finally develop extensive computational experiments on square grids and randomly generated graphs with up to 120 vertices, and a number of components ranging from 2 to 9. We compare the performance of our formulations, all of which use flow-type constraints to model connectivity, against the alternative of using the well known Miller-Tucker-Zemlin constraints, providing a comparative analysis of their performance.

# Bi-Level Service Design Problem considering outsourcing costs

Elena Fernandez,[1] Ivana Ljubić,[2] and Nicolas Zerega[3]

[1]*University of Cadiz, Spain,* elena.fernandez@uca.es

[2]*ESSEC Business School, France,* ivana.ljubic@essec.edu

[3]*University of Cadiz, Spain,* nicolas.zerega@uca.es

In recent years, there has been a significant increase in the outsourcing of various practices. This trend has been particularly prominent in the logistics sector, where it encompasses activities like last-mile delivery and full integration with external operators, including 3-PL logistics partners [1]. Similarly, the airline industry has also experienced outsourcing in processes such as check-in, luggage management, and even cabin crew [3], among others. Notably, there are cases in which flights are outsourced to third-party airlines, by taking advantage of the existing transfer system in major airports [2]. The outsourcing of these processes offers several advantages, including enhanced flexibility and a reduced dependency on hiring and training specialized staff.

Given the increasing prevalence of this trend, it is crucial to study and model this type of situations from a network design perspective to gain a better understanding of how the outsourcing decisions contributes to the overall revenue of a major firm when it faces such a process. It is equally important to consider the companies' viewpoint as they too aim to optimize their revenues.

The objective of this research is to address the transportation of demand between different origins and destinations (commodities) when the major firm, referred to as the Leader, already possesses a hub network of major hubs and chooses to outsource the demands coming from and going to the remaining non-hub locations using third-party companies (carriers) in order to maximize its overall profit. The incoming trips, from a non-hub to a hub node are referred to as first legs, and the outgoing trips, from a hub to a non-hub node, as third legs.

The Leader is responsible for making strategic decisions regarding the commodities that will ultimately be served. The Leader also has to make tactical decisions regarding the assignment of carriers to non-hub origins and destinations. Additionally, the Leader makes operational decisions regarding the routing of served commodities through the hub network. On the other hand, carriers have the authority to determine which commodities can be served by activating their first or third leg, based on the offer and assignment provided by the Leader, taking into account their reservation prices.

We model this problem as a Bi-Level Mixed Integer Non-Linear Programming (MINLP) model, which we subsequently linearize to obtain a Bi-Level Mixed Integer Linear Programming (MILP) formulation. Leveraging the inherent properties associated with the independency of assignments and costs of each carrier, we discretize the outsourcing cost decisions, enabling us to express the model as a Single-Level MILP. The costs and solutions derived from solving this model are shown to be bi-level optimal.

We present and compare several approaches to solve this problem. The first approach distinguishes each type of outsourcing cost based on the carrier, commodity, and hub used for the connection. The second approach aggregates the first and third legs by carriers. The third formulation explicitly focuses on the commodities, while the fourth formulation determines the routing of commodities as an implicit path formulation based on carrier assignments.

Computational results demonstrate the superiority of the implicit paths formulation which is able to solve instances of 200 nodes and 6 carriers to optimality within one hour. These findings provide motivation for studying more complex systems in the future.

# Prize-collecting Location Routing on Capacitated Trees[*]

Elena Fernandez,[1] and M. Munoz-Marquez[2]

[1]*Universidad de Cádiz, Spain,*  elena.fernandez@uca.es

[2]*Universidad de Cádiz, Spain,*  manuel.munoz@uca.es

In this work we study Capacitated Prize-collecting Location Routing Problems (CPLRPs) on trees. Some Lagrangian relaxations are tested and compare. An extensive computational study has been carried out to test the performance of the proposed relaxations.

## 1.     Introduction

In CPLRPs there is a set of users with demand, located at the vertices of a tree, each of which is associated with a profit, and each edge has a cost. To serve the demand a set of routes must be set up, each of them starting and ending at a open node. The activation of a node has an associated cost. In addition, each node has a capacity that must not be exceeded by the demand of the nodes associated with the route that starts from that node.

The objective is to maximize the total net profit, defined as the total income from service to the selected demand vertices, minus the total cost that includes the overall set-up cost of activated facilities plus the routing cost of the edges used in the routes.

## 2.     Results

For the uncapacitated problem, a mathematical programming formulation is presented, which has the integrality property. The formulation models a directed forest where each connected component hosts at least one open facility, which becomes the root of the component.

---

[*]Thanks to everyone...

For the capacitated version, an extensive computational study has been carried out to test the performance of some Lagrangian relaxation.

# Single-allocation hub location with upgraded connections *

Mercedes Landete,[1] Juan M. Muñoz-Ocaña,[2] Antonio M. Rodrígez-Chía,[2] and Francisco Saldanha-da-Gama[3]

[1]*Departamento de Estadística, Matemáticas e Informática, Centro de Investigación Operativa, Universidad Miguel Hernández, Spain,*  landete@umh.es

[2]*Departamento de Estadística e Investigación Operativa, Universidad de Cádiz, Spain,* juanmanuel.munoz@uca.es ,antonio.rodriguezchia@uca.es

[3]*Sheffield University Management School, Sheffield, United Kingdom,* francisco.saldanha-da-gama@sheffield.ac.uk

## 1.      Introduction

In this work we assume that there is a budget that can be invested in improving network connections, understanding that improving a connection can consist of activities as diverse as changing a vehicle for another with less fuel consumption or paying a small fee to avoid traffic jams or even paving a path. In this context, it is possible to solve the single hub location problem at the same time the set of connections to upgrade is decided. Two types of connections are distinguished, the connections between hubs and the connections between spokes and hubs and a quota is established for both values.

A similar problem was analysed in [1] where the authors considered a so-called tree-of-hubs location problem, which is a hub location problem with tree topology for the hub-level network. In this work, we do not impose any hub topology on the hub-level network.

---

*This work was partially supported by the grants PID2021-122344NB-I00, by the Spanish Ministry of Science and Innovation, and PROMETEO/2021/063 by the governments of Spain and the Valencian Community.

# 2.  The problem

We propose two different models for the problem and we conduct several enhancements of both models. The first model is a flow model and the second is a model based on the discrete ordered median problem. Both models allow more than two intermediate hubs for each origin-destination pair since the triangle inequality property can be lost when upgrades are made on connections. The enhancements of the models are based on the theory for binary quadratic programs in [2].

Finally, we show how to adapt the models to the incomplete inter-hub network case and we present an exhaustive computational analysis to compare the efficiency of both formulations.

# Framework for Fairness Analysis in Location Problems

Ivana Ljubić [1], Miguel A. Pozo [2], Justo Puerto [3] and
Alberto Torrejón [4]

[1] *ESSEC Business School of Paris, France*
ljubic@essec.edu

[2] *University of Seville, Spain*
miguelpozo@us.es

[3] *University of Seville, Spain*
puerto@us.es

[4] *University of Seville, Spain*
atorrejon@us.es

Ordered optimization allows a straightforward and flexible generalization for facility location problems since many well-known problems in the literature can be modeled using an ordered formulation. This is the case for median, center or centdian problems, but also, since negative and non-monotonous vectors of weights are allowed, of obnoxious location problems, modeling preferences or other abstract concepts such as, for example, envy. In the context of fairness theory, the ordered optimization can be used to create a common framework for the inclusion and analysis of the concept of equity in location problems, where compensation of allocation costs is based on the fact that a solution that is good for the system does not have to be acceptable for all single parties if their costs to reach the system are too high in comparison to other parties.

In this communication, we present new enhancements for the Discrete Ordered Median Problem (DOMP), a location problem where client-facility allocations are ranked, see [1] & [2]. First, we review the current most competitive formulations in the literature and show how the DOMP can be used for the generalization of many facility location problems. We use a Benders decomposition approach to improve state-of-the-art results for a general criteria selection. We also describe how to include inter-facility connection to the DOMP by means of a minimum spanning tree, in order to present how to apply the DOMP as a framework for fairness analysis in connected facility problems as remark.

# Planar location problems with uncertainty demand points: robustness concepts

Juan A. Mesa,[1] and Anita Schöbel[2]

[1] *Universidad de Sevilla, Spain,*  jmesa@us.es

[2] *Techical University of Kaiserslautern, Germany,*  schoebel@mathematik.uni-kl.de

Robustness was defined in the glossary of the Institute of Electrical and Electronic Engineering (1991) as "the degree to which a system or component can function correctly in the presence of invalid inputs or stressful environmental conditions", whereas the Cambridge English Dictionary as "Applied to a calculation, process, or result if the result is largely independent of certain aspects of the input". In applications, there are many interpretations of robustness depending on the context, type of disruption or perturbation, aims, fields of applications, etc. The concept of robustness can be considered inversely related to that of the vulnerability of a system which has been also defined as the loss of efficiency. However, there are several ways to define the efficiency of a system, and therefore to define vulnerability. Robustness has been confounded with resilience in the sense of the increasing ability to absorb disturbances and perturbations. However, resilience is also interpreted as the rapidity to recover basic functionality from errors, failures, or environmental causes.

Robust Optimization is the branch of Mathematical Optimization that deals with problems where the parameters or input are uncertain and is not known or it is not applicable to this uncertainty a known probability distribution. Some questions arise from this definition. Which are the parameters affected by the uncertainty? To which set do the values of the uncertain parameters belong? How strict should the applied concept be? These and other questions have given rise to different concepts of robustness. Finding robust solutions to an optimization problem is an important issue in practice. In many applications, data and parameters are not precisely known. The uncertainty can have many different reasons, two of which are unknown data (due to measurement errors or some behavior of customers that only can be estimated) or disturbances/perturbations produced by environmental to system effects. In Location Science, weights allocated to demand points, distances from the demand points to facilities, and their position are often not exactly known.

In this paper, we suppose that the coordinates of the demand points are uncertain, but they belong to a given neighborhood of the nominal scenario. We apply several approaches of robust optimization to a broad class of planar location problems. Various concepts on how to define the robustness of an algorithm or of a solution have been suggested, among them, strict [1] , cardinality-constrained [2], adjustable [3], light [4], [5], or recoverable robustness [6]. In this work, we present explicit examples and algorithms for planar location problems and compare their solutions for the different robustness concepts.

# Heuristic segmentations in Electron Tomography using the DOMP [*]

J. M. Muñoz-Ocaña[1], J. Puerto[2] and A. M. Rodríguez-Chía [1]

[1]*Departamento de Estadística e Investigación Operativa, Universidad de Cádiz, Cádiz, Spain*,  juanmanuel.munoz@uca.es, antonio.rodriguezchia@uca.es

[2]*IMUS, Instituto de Matemáticas de la Universidad de Sevilla, Sevilla, Spain.*  puerto@us.es

Today, Scanning Transmission Electron Microscopy plays an important role in designing nanomaterials to be used in the development of different fields, such as green energy sources, catalysis and environmental protection. The Electron Tomography procedure involves a three-stage process: recording the original sample using an electron microscope; reconstructing the object under study from the information provided by the first stage; and segmenting the images before or after they are reconstructed.

The discrete ordered median problem presents an application in electron tomography image segmentations. The adaptability of this problem to the different instances achieves high-quality image segmentations. However, the solutions provided by the discrete ordered median problem for large scale instances are obtained in high computing times. Image size is of great importance in electron tomography experiments, since the larger the image size, the higher the quality of the image. Therefore, applying this problem to images constituted by a large number of intensities could be impractical for electron tomography experiments due to the large number of images to be segmented.

With the goal of reducing the computation times, this work introduces different heuristic procedures to obtain feasible solutions for the ordered median problem that provide high-quality images in low computing times. Moreover, some noticeable improvements of the heuristic techniques are developed taking advantage of the particular versions of the ordered me-dian function that have been proven to be especially suitable in the electron tomography image segmentation.

[*]This work was partially supported by Agencia Estatal de Investigación, Spain and ERDF through projects PID2020-114594GB-{C21, C22} and RED2022-134149-T

# The load–dependent drone general routing problem

Isaac Plana,[1] José María Sanchis,[2] and Paula Segura[3]

[1]*Universitat de València, Spain,*  isaac.plana@uv.es

[2]*Universitat Politècnica de València, Spain,*  jmsanchis@mat.upv.es

[3]*Universitat Politècnica de València, Spain,*  psegmar@upvnet.upv.es

Recent advancements in drone technology have made it possible for these aerial vehicles to have the capability of capturing images and making deliveries during the same flight. This capability has many applications, such as in humanitarian emergencies, where drones can deliver medications to hard–to–reach areas and capture images of disaster–affected zones. However, the payload capacity of drones is still quite limited, and the weight of the cargo being transported significantly impacts energy consumption and flight times.

    In this talk, we present the Load–Dependent drone General Routing Problem (LDdGRP), where a drone must traverse a set of edges and make a certain delivery at some vertices of a graph. The goal is to minimize the total duration of the drone route while accounting for the effect of the load carried by the drone on the traversal time of the edges. We propose a formulation for this problem and conduct an analysis of the associated polyhedron of solutions, while introducing families of valid inequalities to strengthen the formulation. We design a branch–and–cut algorithm to solve the problem and perform extensive computational tests on instances with a maximum of 7 deliveries and 96 edges to be traversed.

# New Formulations and Column Generation Algorithm of the Minimum Normalized Cuts Problem

D. Ponce,[1] J. Puerto[2] and F. Temprano[3]

[1]*Universidad de Sevilla, Spain,*   dponce@us.es

[2]*Universidad de Sevilla, Spain,*   puerto@us.es

[3]*Universidad de Sevilla, Spain,*   ftgarcia@us.es

This paper deals with the k-way normalized cut problem in networks. The normalized cut function was defined to solve some issues concerning the interpretability of the minimum cut problem, which is a classical problem in graph theory whose aim is to provide the bipartition that minimizes the number of edges between nodes from different subsets, applied to partitioning and districting problems. Instead of considering just the number of external edges of each subset, the minimum k-way normalized cut problem tries to minimize the external edge density of each subset of a k-partition, also considering the number of internal edges. In addition, the problem can be extended to a weighted graph in order to minimize the sum of external weight density of the subsets. The minimum k-way normalized cut allows us to locate groups of nodes that accumulate a high internal weight density. Considering that these weights can represent a large number of different interesting parameters, this organization is really useful as facility location planning in order to locate in each of the clusters a facility that accumulates a high weight density. We show several applications of the minimum k-way normalized problem to facility location. We present a methodology using mathematical optimization to provide mixed integer linear programming formulations for the problem. The paper also develops a branch-and-price algorithm for the above mentioned problem which scales better than the compact formulations. Extensive computational experiments assess the usefulness of these methods to solve the k-way normalized cut problem over different location problems on large graphs and random graphs. In addition, all methods have been analysed and studied in order to try to improve them as much as possible.

# Touring polytopes, the weighted region and some related location problems [*]

Justo Puerto [1], Martine Labbé [2], Moises Rodríguez [†]

[1]*IMUS, University of Seville Spain*

[2]*Computer Science Department, Université Libre de Bruxelles, Belgium*

[†]**In Memorian**. *IMUS, University of Seville Spain*

In this paper we address two different related problems. We first study the problem of finding a simple shortest path in a $d$-dimensional real space subdivided in several polyhedra endowed with different $\ell_p$-norms. The second problem that we consider is the continuous single facility location or Weber problem that results in this subdivision of $\ell_p$-normed polyhedra. The first problem is a variant of the weighted region problem, a classical path problem in computational geometry introduced in Mitchell and Papadimitriou (JACM 38(1):18-73, 1991). As done in the literature for other geodesic path problems, we relate its local optimality condition with Snell's law and provide an extension of this law in our framework space. We propose a solution scheme based on the representation of the problem as a mixed-integer second order cone problem (MISOCP) using the $\ell_p$-norm modeling procedure given in Blanco et al. (Comput Optim Appl 58(3):563–595, 2014). We derive two different MISOCPs formulations, theoretically compare the lower bounds provided by their continuous relaxations, and propose a preprocessing scheme to improve their performance. The usefulness of this approach is validated through computational experiments. The formulations provided are flexible since some extensions of the problem can be handled by transforming the input data in a simple way. To solve the second problem, we adapt the solution scheme that we developed for the shortest path problem and validate our methodology with extensive computational experiments.

---

[*]This talk is dedicated to Moises wherever he is.

# The Hampered Travelling Salesman Problem with Neighbourhoods

Justo Puerto [1] and Carlos Valverde[1]

[1]*University of Seville, Spain,*  puerto@us.es

[2]*University of Seville, Spain,*  cvalverde@us.es

This work deals with two different route design problems in a continuous space with neighbours and barriers: the shortest path and the travelling salesman problems with neighbours and barriers. Each one of these two elements, neighbours and barriers, makes the problems harder than their standard counterparts. Therefore, mixing both together results in a new challenging problem that, as far as we know, has not been addressed before but that has applications for inspection and surveillance activities and the delivery industry assuming uniformly distributed demand in some regions.

We provide exact mathematical programming formulations for both problems assuming polygonal barriers and neighbours that are second-order cone (SOC) representable. These hypotheses give rise to mixed integer SOC formulations that we preprocess and strengthen with valid inequalities. The work also reports computational experiments showing that our exact method can solve instances with 75 neighbourhoods and a range between 125-145 barriers.

# Budget constrained cut problems

Justo Puerto[1], José Luis Sainz-Pardo [2]*

[1]*IMUS (Universidad de Sevilla), Spain,*   puerto@us.es

[2]*Centro de Investigación Operativa (Universidad Miguel Hernández de Elche), Spain,*
jlsainz@umh.es

## 1.      Introduction

The minimum and maximum cuts of an undirected edge-weighted graph are classic problems in graph theory. While the Min-Cut Problem can be solved in $P$, the Max-Cut Problem is NP-Complete. Exact and heuristic methods have been developed for solving them. For both problems, we introduce a natural extension in which cutting an edge induces a cost. Our goal is to find a cut such that minimises the sum of the cut weights but, at the same time, it restricts its total cut cost to a given budget. We prove that both restricted problems are NP-Complete and we also study some properties. Finally, we develop exact algorithms to solve both and also a non-exact algorithm for the min-cut case based on a Lagreangean relaxation that provides most of the times optimal solutions. Their performance is reported by an extensive computational experience.

## 2.      Computational Experience

### 2.1      Budget Constrained Min-Cut Problem

Tables 1-3 summarize the computational experience about the Budget Constrained Min-Cut Problem.

| t IP | t Algorithm | # Conc. |
|------|-------------|---------|
| 769.22 | 1.22 | 1538 |

*Table 1.* Instances solved by both exact methods (min-cut)

| # Non-conc. | t Alg. | # Coincs. | % GAP | # Non-sol. |
|-------------|--------|-----------|-------|------------|
| 82 | 1.37 | 20 | 161.58 | 13 |

*Table 2.* Instances non-solved by IP (min-cut)

| t IP | t Lag. | # Non-Opt | % GAP |
|------|--------|-----------|-------|
| 614.16 | 0.03 | 132 | 1.93 |

*Table 3.* Lagrangian relaxation summary

## 2.2    Budget Constrained Max-Cut Problem

Table 4 summarizes the computational experience about the Budget Constrained Min-Cut Problem.

| t LP | t Algorithm | # Conc. |
|------|-------------|---------|
| 207.30 | 65.25 | 1948 |

*Table 4.* Instances solved by both methods (max-cut)

# An exact approach towards solving a hydrogen refuelling network design problem with pipelines

K.D. Searle,[1][*]C. Searle,[2] and J. Kalcsics[1]

[1]*School of Mathematics and Maxwell Institute for Mathematical Sciences, University of Edinburgh*

[2]*Centre for Logistics and Sustainability, Edinburgh Business School, Heriot-Watt University*

In this presentation I will present a network design model which is employed to obtain the minimum set-up and operational cost hydrogen refuelling infrastructure. The network design model takes into consideration both facility location and hydrogen supply decisions. More specifically, the model must determine where to locate hydrogen refuelling stations and how the hydrogen must be supplied to each refuelling station. For the supply of hydrogen we consider local hydrogen supply and off-site hydrogen supply. In the latter case the hydrogen must be distributed to the refuelling stations by means of tube trailers or a hydrogen pipeline. The three supply modes (localised production, tube trailer, and pipelines), result in model solutions that take on a forest-stars-points topology. More specifically, the hydrogen refuelling stations connected to the hydrogen pipeline will resemble a forest of directed trees rooted at hydrogen pipeline supply points, while the hydrogen refuelling stations supplied by tube trailers will resemble a set of stars centered around the centralised production facilities and, finally, the hydrogen refuelling stations which have localised hydrogen production will resemble a set of isolated points. The difficulty of solving the network design problem arises from the hydrogen pipeline routing decisions which will be the focus of the presentation. More specifically, I will present three different formulations and two different cutting plane algorithms to solve this problem. We then compare each of the formulations and solution algorithms in a preliminary computational study where we found that the classical flow formulation with a set of valid inequalities is the most efficient. Finally, I will present a real-life case study in the north of England where we found that allowing hydrogen to be delivered by pipelines made a significant improvement in the objective function.

---
[*]Corresponding author,  kd.searle@ed.ac.uk

# A novel integrated approach for locating agricultural sensors based on the zoning and p-median problems

Tiare Torres[1*], Víctor M. Albornoz[1], and Rodrigo Ortega[2]

[1]*Universidad Técnica Federico Santa María, Dpto. de Industrias,Santiago, Chile,*

[2]*Universidad Técnica Federico Santa María, Dpto. de Ing. Comercial,Santiago, Chile,*

[*]   tiare.torres@sansano.usm.cl

Precision Agriculture (PA) is a very relevant field of study nowadays. Its objective is to increase efficiency, yields, productivity and, at the same time, reduce the environmental impact of usual agronomic practices in the agricultural industry through the use of technology and data analysis.

Due to the benefits that the use of PA brings, many farmers decide to implement them, but due to the costs, they can only have access to a limited number of devices. Therefore, in this contribution we will develop an integrated approach for the management zone delineation problem and the discrete location of sensors problem to increase the representativeness of nonhomogeneous soil properties.

The proposed model is solved using the integer programming solver Gurobi in Julia JuMP and computational results from a set of instances are presented to show the impact of the adopted methodology.

## 1.      Problem Description

The first step in PA implementation is data collection. In this regards, satellite images, drones, and fixed sensors on the ground are used to record different kinds of data. Specifically, there is a wide variety of sensors, with specific characteristics and an extensive price range.

One type of fixed sensors on the ground works only for point measurement, and can be located with some depth. Point measurement means that the sensor cannot collect data in a radius around it. A common practice in this field is to take the data and interpolate for the whole area, obtaining continuous data from a discrete parameter.
Sometimes, due to the cost of equipments, small farmers can only access a limited number of them. A smaller sample size, makes data interpolation less reliable causing subsequent analyses to be misleading. For this reason, it is interesting to determine where a limited number of sensors can be located to obtain the most representative sample.

## 2. Methodology and Results

The zoning problem is a PA problem that its aim is to divide a field into homogeneous management zones. The binary integer program in [1] provides rectangular management zones and uses a fixed level of relative variance [2] to guarantee the homogeneity.

These management zones can be represented as a node in a network and a p-median location model [3] allows to decide where locate a limited number of sensors. The objective of the model is minimize the weighted distance from each management zone to the zones that has the device, and the weight is the variance of each zone.

The problem can be solved considering a hierarchical approach, where two models are solved sequentially, one for the zoning problem and then another one for the location problem, using the resulting partition of the first one. On the contrary, the *integrated model* solves both problems simultaneously, i.e. the model find the optimal partition of the field and in which management zones to locate the devices.

The two approaches are compared using different instances considering the size of the field, the level of homogeneity required and the number of sensors available. The Integrated model shows better solutions than the hierarchical approach in most instances, but it presents a higher computational cost.

# Author Index

# S